\documentclass[journal]{IEEEtran}
\usepackage{cite}
\usepackage{multirow}
\usepackage{amsmath}
\usepackage{graphicx}
\usepackage{booktabs}
\usepackage{subfigure}
\usepackage{stfloats}
\usepackage{makecell}
\usepackage{color}
\usepackage{amssymb}
\usepackage{pifont}
\usepackage[ruled,linesnumbered]{algorithm2e}
\usepackage[table,xcdraw]{xcolor}
\usepackage[pagebackref=false,breaklinks=true,letterpaper=true,colorlinks,bookmarks=false]{hyperref}
\ifCLASSINFOpdf
\else
\fi

\begin{document}
%
\title{Dual-Stage Approach Toward Hyperspectral Image Super-Resolution}
%
%
%

\author{Qiang~Li,~\IEEEmembership{Student Member,~IEEE,}
        Yuan~Yuan,~\IEEEmembership{Senior Member,~IEEE,}
        ~Xiuping~Jia, ~\IEEEmembership{Fellow,~IEEE,}
        and~Qi~Wang,~\IEEEmembership{Senior Member,~IEEE}

\thanks{
	
	Qiang Li, Yuan Yuan, and Qi Wang are with the  School of Artificial Intelligence, Optics and Electronics (iOPEN), 	Northwestern Polytechnical University, Xi'an 710072, P. R. China. Qiang Li  is also with the School of Computer Science, Northwestern Polytechnical University, Xi'an 710072, P.R. China (e-mail: liqmges@gmail.com, y.yuan1.ieee@gmail.com, crabwq@gmail.com).  
		
%
	
	Xiuping Jia is with the School of Engineering and Information Technology,
	University of New South Wales, Canberra, ACT 2612, Australia (e-mail:
	x.jia@adfa.edu.au)
	 }}
 
\markboth{}%
{Li \MakeLowercase{\textit{et al.}}: Dual-Stage Approach Toward Hyperspectral Image Super-Resolution}

\maketitle

\begin{abstract}
	
Hyperspectral image produces high spectral resolution  at the sacrifice of spatial resolution.  Without reducing the spectral resolution, improving the resolution in the spatial domain is a  very challenging problem. Motivated by the discovery that hyperspectral image exhibits  high similarity between adjacent bands in a large spectral range, in this paper, we explore a new structure for hyperspectral image super-resolution (DualSR), leading to a dual-stage design, i.e., coarse stage and fine stage.  In coarse stage, five bands with high similarity in a certain spectral range are divided into three groups, and the current band is guided to study the potential knowledge. Under the action of alternative spectral fusion mechanism, the coarse SR image is super-resolved in  band-by-band. In order to build model from a global perspective,  an enhanced back-projection method via spectral angle constraint is developed in fine stage to learn the content of spatial-spectral consistency, dramatically improving  the performance gain.   Extensive experiments demonstrate the effectiveness of the proposed coarse stage and fine stage. Besides, our network produces state-of-the-art results against existing works in terms of spatial reconstruction and spectral fidelity.   Our code
is publicly available at \href{https://github.com/qianngli/DualSR}{https://github.com/qianngli/DualSR}.

\end{abstract}


\IEEEpeerreviewmaketitle

\section{Introduction}

\IEEEPARstart{H}{yperspectal} image  is widely utilized in various fields because of its rich spectral information. Generally, several discriminant bands in hyperspectral image are selected for subsequent analysis according to the spectral characteristics of the objects of interest. \cite{bajwa2004hyperspectral}. However, hyperspectral image produces high spectral resolution  at the sacrifice of spatial resolution \cite{dian2017hyper, kawa2011hyper}, which cannot meet the requirements of some scene applications. Considering this dilemma, the researchers propose hyperspectral image super-resolution (SR). Without reducing the number of bands, hyperspectral image SR aims to find a high-resolution (HR) image with better visual quality and refined details from counterpart low-resolution (LR) version in spatial domain. SR  is a greatly challenging task in  computer vision, because it is an ill-posed inverse problem, i.e. there are multiple solutions for the same LR image.

Currently, the existing techniques  can be divided into two categories, including single hyperspectral image SR without any auxiliary information \cite{li2020exp, jiang2019learning, li2019dual, liu2021a} and single hyperspectral image SR using a given HR multispectral image \cite{zhang2020unsiper, pan2019multi, xie2019multi, qu2018unsuper, dian2017hyper}. In our paper, we employ the former to build the model.  As for this type, previous approaches are roughly classified into three groups. They are \textit{2D convolutional neural network (CNN)}\cite{hu2020hyperspectral, jiang2019learning}, \textit{3D CNN} \cite{li2019hyperspectral,jiang2019learning, Mei2017Hyperspectral, yang2020hb}, and \textit{2D/3D CNN} \cite{li2020exp, Wang2020sfcsr}.
With respect to the first group, it usually regards  each band in hyperspectral image as an image to design the model using 2D convolution. This way is similar to natural image SR. Thus, the approaches for natural image SR \cite{fang2020soft, Zhang2020coar} can directly be applied to this field.  As  mentioned above, the spectrum curve of the target is usually used to analyze whether there is a corresponding target in practical applications. This means that the reconstructed spectral curve maintains the original curve characteristics as much as possible, so that the target can be found according to the spectral curve. This is also one of key differences  between hyperspectral image SR and  natural image SR. Nevertheless, the first group solely investigates spatial knowledge, which results in inferior spectral fidelity. 

\begin{figure}[t]
	\centering
	\includegraphics[height=4.6cm,width=0.41\textwidth]{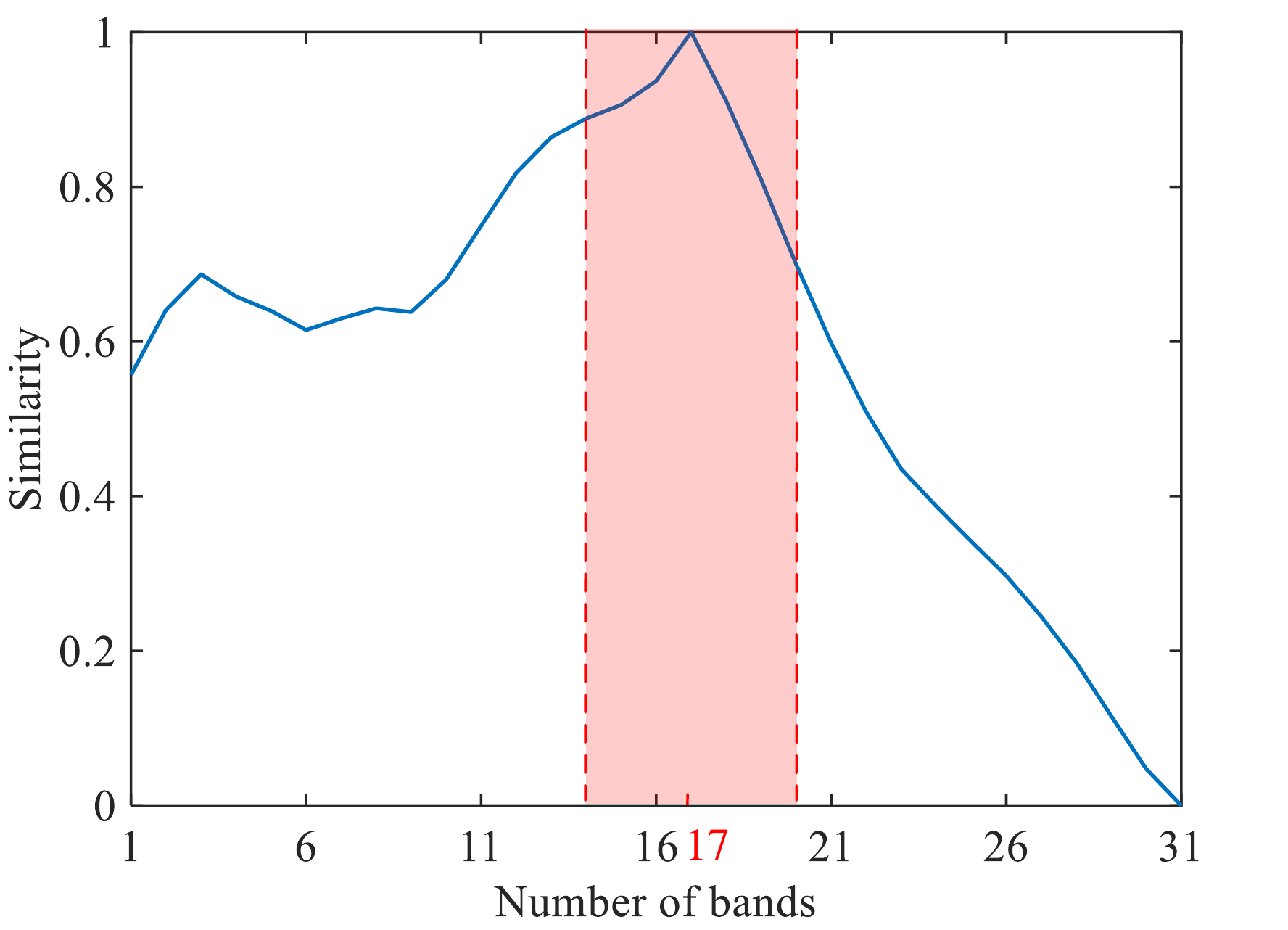}
	\caption{Illustration of the existence of high similarity in adjacent bands.  We provide the similarity values  between the 17\textit{-th } band and the other bands. The similarity value here is calculated by the reciprocal of Euclidean distance.}
     \vspace{-0.3cm}
	\label{fig:simi}
\end{figure}

Hyperspectral image shows rich spectral knowledge, which is helpful to enhance the performance of super-resolved image in spatial aspect. Considering this fact, various hyperspectral image SR algorithms based on 3D convolution are developed \cite{li2019hyperspectral,jiang2019learning, Mei2017Hyperspectral, li2019dual, Mei2020spa}. By contrast, the performance of these studies is superior for the first kind of method, which benefits from the exploration of spectral dimension. Among these algorithms, the regular 3D convolution has been  abandoned due to the large number of parameters and turns to separable 3D convolution \cite{li2019hyperspectral, li2019dual, Zolfaghari2018eco}. As for hyperspectral image SR, using the abundant spectra mainly attempts to enhance spatial performance. In this respect, we should treat the analysis of spectral and spatial domain differently. However, these works ignore this point.



To cope with the above problem, the algorithms using 2D/3D convolution are proposed \cite{Li2020Mixed, li2020exp, Wang2020sfcsr}. In contradiction to existing works, this type of approach adopts mixed 2D/3D convolution to establish the model. On the premise of obtaining spectral information, more 2D convolutions are attached with model to heighten the feature learning ability in space. Meanwhile, it dramatically decreases the parameters.  The above approaches fixedly inputs all the bands of the hyperspectral image into the model. Since the hyperspectral image contains dozens or even hundreds of bands, this inevitably requires more GPU memory. Inspired by  high similarity between adjacent bands \cite{Yuan1017hyperspectral, wang2020hyperspectral}, Wang \textit{et al.}  \cite{Wang2020sfcsr} create a novel dual-channel structure to establish network. Impressively, unlike existing methods, it integrates the information of single LR band and two adjacent LR bands to achieve super-resolved band. At present, there is extremely little research using this novel input mode.


 
In fact, within a certain spectral range, relatively distant bands can also explicitly assist the current band to reconstruction, because these bands are also similar, but  the similarity is relatively small (see Fig. \ref{fig:simi}). If more adjacent bands  within a relatively large spectral range are  utilized, it is  beneficial to supplement the missing knowledge during   the reconstruction of current band. Therefore, the key problem is \textit{how to effectively use the adjacent bands to boost performance}. Motivated by these discoveries, we propose a dual-stage approach toward hyperspectral image SR (DualSR). In coarse stage,  we first select current band and its four bands with high similarity  within a certain spectral range. Then these bands are divided into three groups to study the potential knowledge for current band. To effectively utilize these bands,  an adjacent spectral fusion mechanism is developed to generate complementary  content from intra/inter-groups, improving representation learning. Then each initial SR band in hyperspecral image is obtained by recurrent manner. Due to the lack of exploration of a larger perspective in spectral dimension, an enhanced back-projection  under spectral angle constraint is designed to further refine result in fine stage. Experiments demonstrate that the proposed dual-stage SR algorithm outperforms the existing works.

In summary, the contributions of this paper are four-fold:

$\bullet$ A novel dual-stage structure to hyperspectral image SR is proposed. In coarse stage, the coarse SR image is  obtained in band-by-band using supervised manner, which achieves  excellent performance. In fine stage,  the coarse result is refined in a global perspective using unsupervised manner, dramatically improving  the performance gain. 

$\bullet$ Neighboring band partition (NBP) is  presented  to group  five bands with high similarity in a certain spectral range into three groups, which modulates the current band to study potential knowledge.

$\bullet$ An adjacent spectral fusion mechanism based on intra/inter-groups is proposed, which generates complementary content from adjacent bands. It  enables the model to learn more missing details in process of single band reconstruction.

$\bullet$ An enhanced back-projection method via spectral angle constraint is developed. Under the condition of spectral angle constraint, the result  is optimized by calculating the reconstruction error. It belongs to a plug-and-play method, which is suitable for any hyperspectral image SR algorithms.

This paper is an extension of our previous work on hyperspectral image SR \cite{Li2021hyper} in the IEEE International Conference on Acoustic, Speech, and Signal Processing. Compared with the conference version, this paper has some extensions as follows:

$\bullet$ \textbf{Methodology:} After obtaining super-resolved image in coarse stage, we add a plug-and-play method as post-processing to further prompt the performance by unsupervised manner, outperforming others by a non-negligible margin.

$\bullet$ \textbf{Experiments:} More experiments are conducted to verify the effectiveness of the proposed model on two datasets, such as more scale factors, improvement of performance, generalizability to various blur kernels, etc. Note that since we modify the parameters of the model in coarse stage, such as  batchsize, rate learning, etc, the obtained results  are higher than that of the conference version.

The rest of this paper is organized as follows. Section \ref{realted} reviews  related hyperspectral image SR algorithms. Section \ref{pm} mainly describes the details in terms of coarse stage and fine stage. The experimental results are analyzed and discussed in Section \ref{es}. Finally, this work is summarized in Section \ref{con}.

\section{Related Works}
\label{realted}
While hyperspectral image SR has an extensive history, in this section, we mainly describe the methods based on CNN, which involves the input mode, the type of convolution used, and some two-stage methods.

\subsection{Input Mode}
\begin{figure*}
	\subfigure[SSRSR]{
		\begin{minipage}[]{0.18\linewidth}
			\centering
			\includegraphics[width=\linewidth]{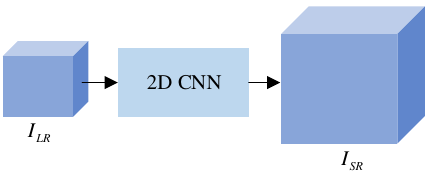}
			\vspace{-0.18cm}
			\label{fig:2dcnn} 	
		\end{minipage}
	} \hspace{0.6cm}
	\subfigure[3D-FCNN]{
		\begin{minipage}[]{0.25\linewidth}
			\centering
			\includegraphics[width=\linewidth]{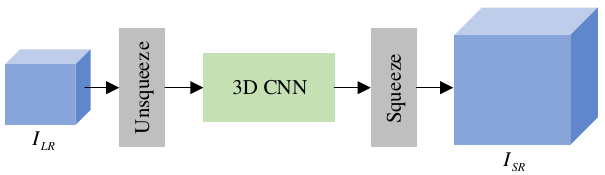}
			\vspace{-0.18cm}
			\label{fig:3dcnn} 
		\end{minipage}
	}\hspace{0.8cm}
	\subfigure[ERCSR]{  
		\begin{minipage}[]{0.32\linewidth}
			\centering
			\includegraphics[width=\linewidth]{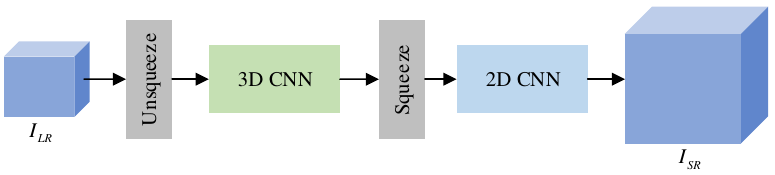}
			\label{fig:3_2dcnn}
			\vspace{-0.25cm} 
		\end{minipage}
	}   
	\hfill
	\subfigure[SFCSR]{
		\begin{minipage}[]{0.39\linewidth}
			\centering
			\includegraphics[width=\linewidth]{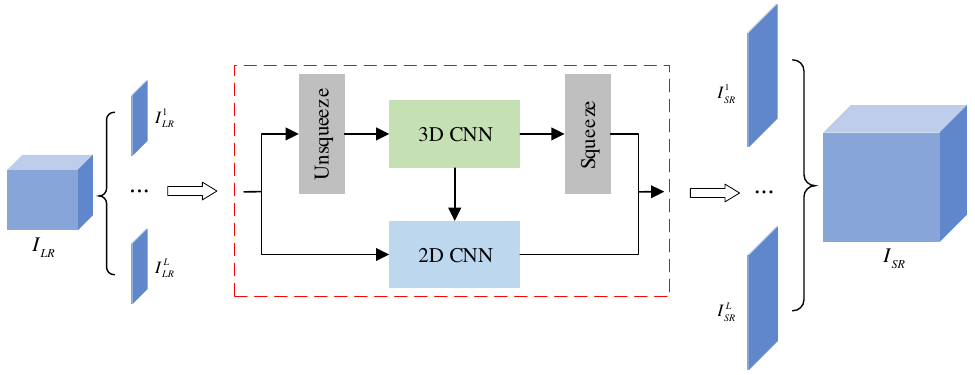}
			\vspace{-0.25cm}
			\label{fig:2_3dcnn} 
		\end{minipage}
	}
	\hspace{0.1cm}
	\subfigure[DualSR (Ours)]{
		\begin{minipage}[]{0.56\linewidth}
			\centering
			\includegraphics[width=\linewidth]{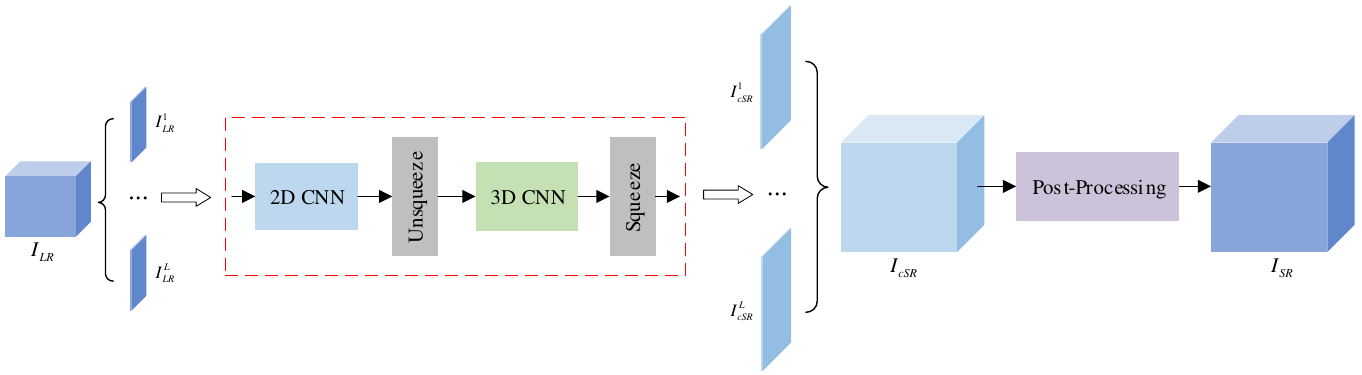}
			\label{fig:ours}
			\vspace{-0.25cm} 
		\end{minipage}
	}	
	\caption{Simplified structure of several methods. $I_{LR}$ denotes LR hyperspectral image. $I_{SR}$ is super-resolved hyperspectral image. $I^i_{LR}$ ($1 \le i \le L$) represents $i$\textit{-th} band of LR hyperspectral image, where $L$ is the total number of bands in hyperspectral image cube.   $I^i_{cSR}$ ($1 \le i \le L$) represents $i$\textit{-th} band of  coarse SR hyperspectral image.}
	\label{fig:method}
\end{figure*}

Unlike natural images, hyperspectral images have multiple bands, which can effectively identify objects. As for hyperspectral image SR,  whether CNN is built with 2D or 3D convolution,  the existing methods \cite{liu2021a, 9097432, Mei2017Hyperspectral, li2020exp, Wei2020deep} are mainly to input all bands into the model for study. The simplified structures are shown in  Figs. \ref{fig:2dcnn}-\ref{fig:3_2dcnn}. It obviously requires more GPU memory than the RGB image SR task.  Under the condition of limited hardware resources, GPU memory is still a major constraint to build a deep network.

To alleviate this trouble, a natural improvement is to change input mode. Concretely, all the bands of hyperspectral image are input into the model in band-by-band. Typical work is SFCSR \cite{Wang2020sfcsr}. The  structure of its network is displayed in Fig. \ref{fig:2_3dcnn}. Inspired by this new structure, in our paper, we adopt this manner to design the backbone in coarse stage.

\label{rw}

\subsection{CNN Using Different Convolutions} 
In this section, we introduce several representative works using different types of convolution.

(a) \textbf{SSPSR:} For hyperspectral image SR and natural image SR,  there is no difference except for the type of input image. Therefore, Jiang \textit{et al.}  \cite{9097432}  refer to single RGB image SR approaches, and introduce spatial-spectral prior network using 2D convolution to design model, whose simplified flowchart is shown in Fig. \ref{fig:2dcnn}. Since  only 2D convolution cannot discover spectral knowledge, this type of method is hardly employed.

(b) \textbf{3D-FCNN:} The regular 3D convolution is first utilized to explore both spatial context and spectral correlation in neighboring bands \cite{Mei2017Hyperspectral}.  The structure is shown in Fig. \ref{fig:3dcnn}.  As we mentioned previously, the regular 3D convolution  produces large parameters, thus making the network not easy to learn. To relieve this drawback, further improvements include  the use of separable 3D convolution \cite{Mei2020spa, jiang2019learning, li2019hyperspectral, li2019dual} and reducing the number of 3D convolution \cite{li2020exp, Li2020Mixed}.

(c) \textbf{ERCSR:} Li \textit{et al.} state that MCNet \cite{Li2020Mixed} designs the main modules in parallel, which does not effectively combine 2D/3D convolution, resulting in structural redundancy.  For this reason, a novel alternate structure with 2D/3D convolutions is developed \cite{li2020exp} (see Fig. \ref{fig:3_2dcnn}). It not only solves module redundancy.  Besides, it also improves the representation learning in spatial domain. Currently, ERCSR is by far the best performance among the existing techniques.

(d) \textbf{SFCSR:}  The approach \cite{Wang2020sfcsr} first applies novel input mode to analyze the information from both  single band and its two adjacent bands (see Fig. \ref{fig:2_3dcnn}).  Although it  availably saves memory footprint, it does not take into account more neighboring bands.  Importantly, such  single band SR inevitably leads to serious spectral distortion during recurrent strategy.  Furthermore, the recurrent mode ignores the overall learning and optimization of hyperspectral image. Considering this issue, after hyperspectral image is reconstructed in  band-by-band, we add an enhanced  back-projection strategy as post-processing to further optimize result. The simplified  overview is shown in Fig. \ref{fig:ours}.  The strategy  refines the result globally after hyperspectral image is reconstructed in  band-by-band. It can maintain the spectral curve well. Meanwhile, the performance of other metrics has also been improved to a certain extent.

\subsection{Two-Stage Methods}
At present, there are few hyperspectral image SR models based on two-stage strategy. Now, we review existing coarse-to-fine architectures. For instance, Yuan \textit{et al.} \cite{yuan2017hyperspectral} propose a framework by transfer learning from natural image domain to hyperspectral image domain. Its aim is to learn the common properties of the two modes. Then collaborative nonnegative matrix factorization is introduce to enforce collaborations between the observed LR hyperspectral image and the transferred HR hyperspectral image. Although it can refine  transferred results, it requires many iterations and is very time consuming to calculate.  Wang \textit{et al.} \cite{wang2021hyperspectral} first develop a recurrent feedback network to generate each band in hyperspectral image. Then the  Pseudo-3D convolution is utilized to further optimize the initial result by supervised manner. Zhang \textit{et al.} \cite{zhang2020unsupervised} propose a coarse-to-fine scheme, including fusion module and  adaptation module. They are utilized to respectively general image prior and specific hyperspectral image. The adaptive module here is made up of multiple convolutional layers and requires many iterations to refine the results.  Different from these works, to obtain better super-resolved result, in our paper, an enhanced back-projection method is designed to optimize by unsupervised manner. Importantly, the proposed approach does require iteration.

\section{Proposed Method}
\label{pm}
\begin{figure}[tp]
	\centering
	\includegraphics[height= 4.2cm, width=0.45\textwidth]{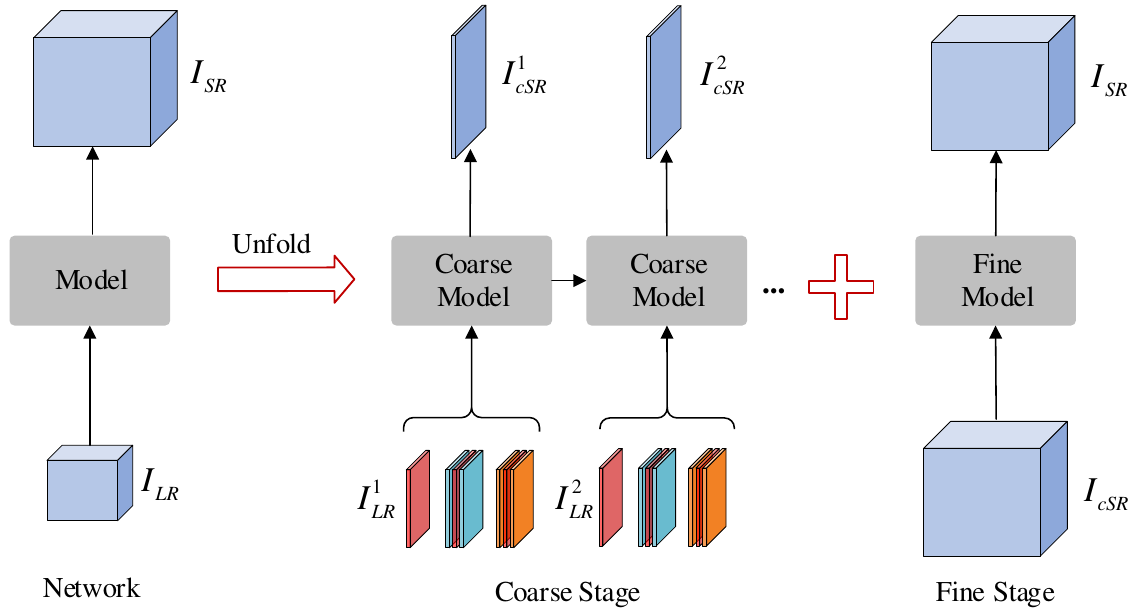}
	\caption{Overview  of the proposed DualSR for hyperspectral image SR.}
	\label{fig:flowchart}
	\vspace{-0.1cm}
\end{figure}

The overall scheme of our proposed DualSR is shown in Fig. \ref{fig:flowchart}. As depicted, our method consists of two steps: coarse stage and fine stage.  In coarse stage, the coarse hyperspectral image is restored in band-by-band using supervised way. Note that the method is our conference version  \cite{Li2021hyper}.  For ease of description, the method  is named CoarSR.  During this process, the current band  is assisted to enhance the exploration ability through four adjacent bands. After obtaining coarse result,  we adopt unsupervised manner in fine stage to globally learn the information, which further optimizes result.
\subsection{Coarse Stage with Supervised Manner}
\begin{figure*}[tp]
	\centering
	\includegraphics[height= 4.2cm, width=0.94\textwidth]{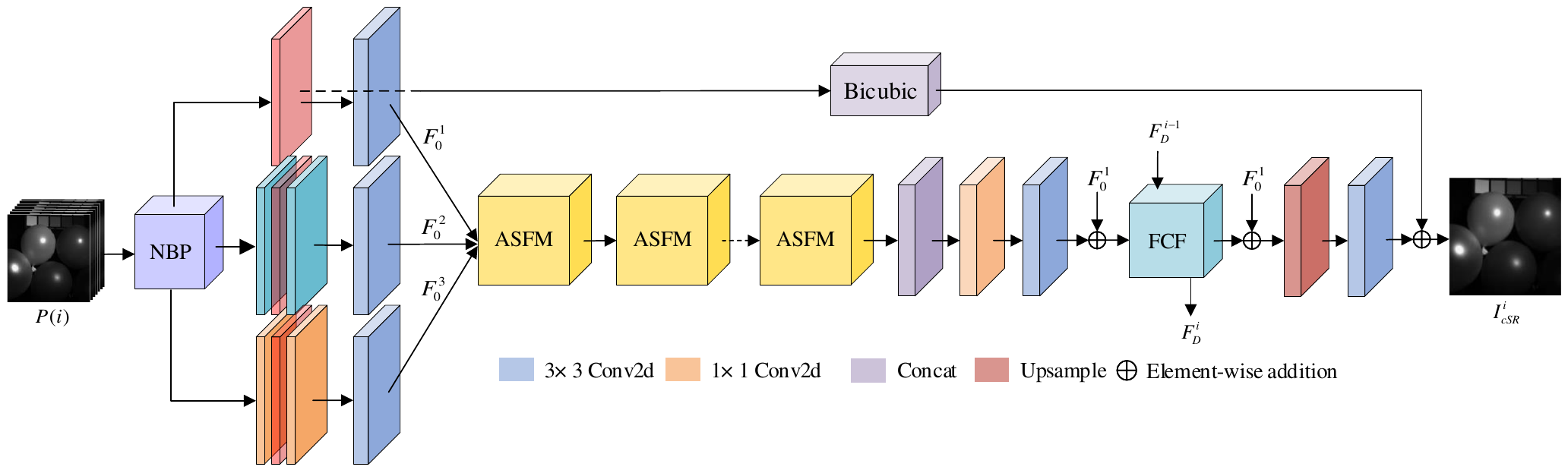}
	\caption{Overview  of the proposed CoarSR for hyperspectral image SR in coarse stage.}
	\label{fig:coarse_flowchart}
\end{figure*}

\vspace{-0.1cm}
\label{cs}
Previous works \cite{jiang2019learning, Li2020Mixed, hu2020hyperspectral} almost adopt this way to build model, i.e.,  all the bands are processed synchronously. Actually, it is not necessary because the reconstruction of the current band is mainly influenced by adjacent bands. Inspired by the pipeline in \cite{Wang2020sfcsr}, we propose a novel structure for hyperspectral image SR via adjacent spectral fusion, whose flowchart is shown in Fig. \ref{fig:coarse_flowchart}. The overview of coarse stage mainly covers three modules, involving  neighboring band partition (NBP), adjacent spectral fusion mechanism (ASFM), and feature context fusion (FCF). 
\subsubsection{Neighboring Band Partition}
SFCSR employs two adjacent bands to discuss, where non-adjacent bands relative to the current band position are not considered.   Fig. \ref{fig:simi} shows that these bands are also greatly similar within a certain spectral range. In fact, relatively distant bands can also explicitly serve the current band.  Therefore, our method adopts single band $I^i_{LR}$ and its four adjacent bands for analysis, i.e.,  
\begin{equation}
P(i) = \left\{ \begin{array}{l}\vspace{0.8ex} 
\left[ {I_{LR}^1,I_{LR}^2,I_{LR}^3,I_{LR}^4,I_{LR}^5} \right],{\kern 20pt}  1\le i < 3\\ \vspace{0.8ex} 
\left[ {I_{LR}^{i-2},I_{LR}^{i-1},I_{LR}^{i},I_{LR}^{i+1},I_{LR}^{i+2}} \right],{\kern 10pt}  3\le i \le L - 3.\\
\left[ {I_{LR}^{L-4},I_{LR}^{L-3},I_{LR}^{L-2},I_{LR}^{L-1},I_{LR}^L} \right],{\kern 1pt}  L - 3 < i \le L
\end{array} \right.
\end{equation}

To assist current band through neighboring bands, in our paper, according to the similarity with the current band, the five bands are partitioned into three groups by neighboring band partition (NBP). As for the case  $3\le i \le L - 3$, it  can be denoted as 
\begin{equation}
\label{group1}
{f_{BP}}(P(i)) = \left\{ \begin{array}{l}\vspace{0.8ex}
I_{LR}^i,\\ \vspace{0.8ex}
\left[I_{LR}^{i-1},I_{LR}^i,I_{LR}^{i+1}\right],{\kern 1pt} 3\le i \le L - 3.\\ 
\left[I_{LR}^{i-2},I_{LR}^i,I_{LR}^{i+2}\right],
\end{array} \right.
\end{equation}
With respect to those bands with larger or smaller index, this is different from the above case.  They are partitioned  by 
\begin{equation}
\label{group2}
{f_{BP}}(P(i)) = \left\{ \begin{array}{l}\vspace{0.8ex}
I_{LR}^i,\\ \vspace{0.8ex}
\left[I_{LR}^1,I_{LR}^2,I_{LR}^3\right],{\kern 1pt} 1\le i < 3,\\ 
\left[I_{LR}^4,I_{LR}^i,I_{LR}^5\right],
\end{array} \right.
\end{equation}
\begin{equation}
\label{group3}
{f_{BP}}(P(i)) = \left\{ \begin{array}{l}\vspace{0.8ex}
I_{LR}^i,\\ \vspace{0.8ex}
\left[I_{LR}^{L-2},I_{LR}^{L-1},I_{LR}^{L}\right],{\kern 1pt} 	L-3< i \le L.\\ 
\left[I_{LR}^{L-3},I_{LR}^i,I_{LR}^{L-4}\right],
\end{array} \right.
\end{equation}

Note that  the current band $I_{LR}^{i}$ appears in each group. 

\subsubsection{Adjacent Spectral Fusion Mechanism}
\begin{figure}[tp]
	\centering
	\includegraphics[height= 4.2cm, width=0.49\textwidth]{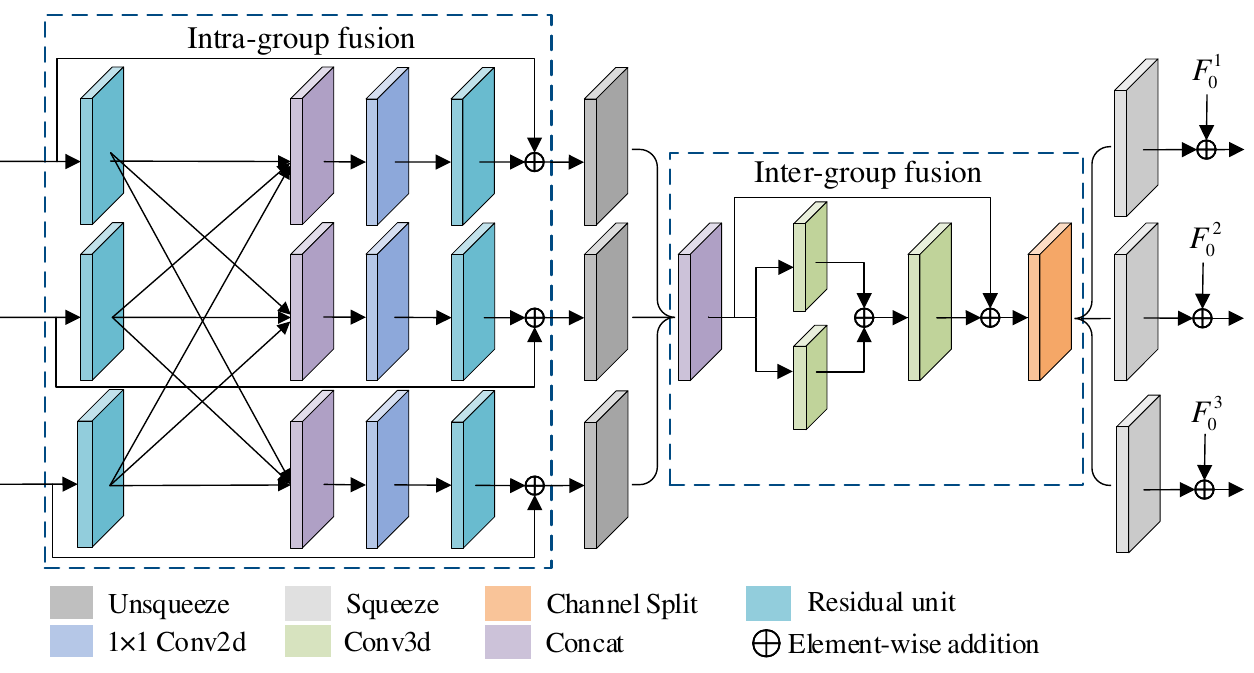}
	\caption{Adjacent spectral fusion mechanism.}
	\label{fig:asfm}
\end{figure}

Adjacent bands not only exhibit high similarity, but also provide the difference in diverse  texture in some regions. It implies the missing information in the current band can be obtained from other bands. Existing works \cite{Li2020al, li2020exp, hu2021hyperspectral} apply small convolution kernel  directly to hyperspectral image. Such the features of non-adjacent bands cannot be  extracted, which hinders the borrowing of complementary content from relatively distant adjacent bands. Considering this issue, an alternative spectral fusion mechanism  (ASFM) is developed to modulate the network to yield more complementary information from intra/inter-groups, which is depicted in Fig. \ref{fig:asfm}.

\textbf{Intra-group Fusion:}
After partitioning these bands, the initial features $F_0^1$, $F_0^2$, $F_0^3$ from three groups are obtained by 2D convolution layer. To sufficiently explore the  potential information in each group, three parallel branches are performed to separately map the three inputs into deep features. For simplicity, we take the first subnet as an example to describe in details.  Since this module only depends on spatial content,  it is handled in much the same way as natural image SR. Therefore, to learn the spatial features of these bands, we employ the main residual module commonly used in natural image SR algorithms  \cite{Anwar2020hyper, hu2020chan} for reference to construct  residual unit.  Concretely, the residual unit in the intra-group part is composed of two identical blocks. Each block contains  two convolution layers with kernel $3 \times 3$,  ReLU function, and skip connection. Moreover, to borrow complementary information within different groups, the local features from other two groups are  concatenated in current group to fuse. Under the action of  intra-group fusion, the features of current band $I_{LR}^i$ in the spatial domain are fully  developed.

\textbf{Inter-group Fusion:}
Only applying 2D convolution in intra-group fusion part cannot study content except for spatial dimension. Hence, to capture spectral features across different groups, we exploit 3D convolution to integrate those through inter-group fusion scheme. However, the direct use of regular 3D convolution significantly produces a large number of network parameters. For this reason, two convolution layers with kernels, i.e., $3\times1\times1$ and $1\times 3\times 3$,  are attached  with network to investigate  spectral and spatial features, respectively. Then,  the information between two domains is availably fused through an addition operation. With respect to hyperspectral image SR, its aim is to heighten spatial resolution while obtaining high spectral fidelity. To achieve this end, the fusion is followed by the convolution layer with kernel $1\times 3\times 3$ to further enhance the learning ability in spatial domain.  By doing so, the inter-group fusion part aggregates the information between different groups. It  realizes the knowledge complementarity across bands in the spectral and spatial domain.

\subsubsection{Feature Context Fusion}
In the process of SR bands, we only pay attention to the knowledge of multiple adjacent bands, i.e., spectral context. Actually, we can also use intermediate features to construct feature contexts because adjacent bands are similar.  At present, feature context has been successfully applied in \cite{Wang2020sfcsr}, which has shown to help model representation learning.  Since our network  is implemented through recurrent strategy in coarse stage, we also adopt this manner to achieve information interaction between bands, namely feature context fusion (FCF).

Let $F_D^i$ and $F_D^{i-1}$ denote  the generated intermediate features  for current bands $i$ and previous band $i-1$. To fuse both features, the concatenation operation is introduced to dynamically generate fusion data through weights $w_1$ and $w_2$.  Then, it is followed by convolution layer with kernel $1 \times 1$  to reduce channels, i.e.,  
\begin{equation}
F = Conv([w_1*F_D^i, w_2*F_D^{i-1}]).	
\end{equation}
Note that FCF is not performed for first band reconstruction. Intuitively, this mechanism is very similar to RNN \cite{mikolov2011extensions}, except that the input band varies with the index increment. 


%
\begin{figure}[tp]
	\centering
	\includegraphics[height= 4.0cm, width=0.48\textwidth]{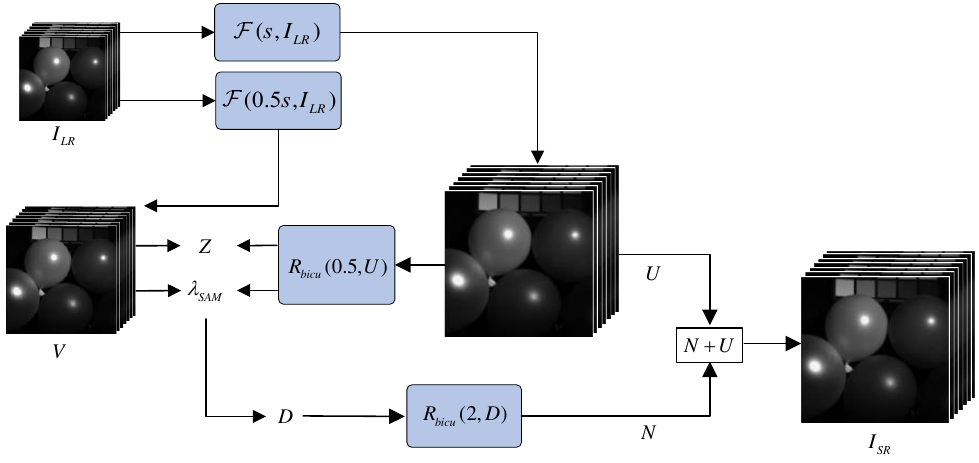}
	\caption{Enhanced back-projection method via spectral angle constraint.}
	\label{fig:fine_flowchart}
\end{figure}
\subsection{Fine Stage Using Unsupervised Manner}
Despite our method utilizes five bands to restore  single band  in coarse stage, it still lacks information exploration of a larger perspective in spectral dimension. As a result, the super-resolved hyperspectral image inevitably suffers serious spectral distortion  by such single band SR. A natural way is to input all bands into the model to optimize the results by supervised manner. Nevertheless, this approach violates the original intention. Considering this limitation, we apply unsupervised manner to further refine initial results.  

Back-projection \cite{Irani1991} optimizes the reconstruction error through an efficient iterative strategy. For this algorithm, it usually utilizes multiple upsampling descriptors to upsample LR image and iteratively calculate the reconstruction error. Currently, back-projection is widely introduced in natural image SR \cite{Yang2010image, Haris2018deep, Haris2019rec}, which  has been proved to develop  the quality of SR image. Yang \textit{et al.} \cite{Yang2010image} employ back-projection to refine results produced according to coupled dictionaries through unsupervised strategy.  However, the initialization which leads to an optimal solution remains unknown. The main reason is that the algorithm involves predefined hyperparameters, such as number of iteration and convolution kernel. 
\begin{algorithm}
	\caption{Dual-stage hyperspectral image SR algorithm (DualSR)}\label{algorithm}
	\KwIn{Hyperspectral image dataset containg LR-HR pair, scale factor $s$}
	\KwOut{Super-resolved hyperspectral image $I_{SR}$}
	Randomly  initialize coarse model parameters $\theta$ \;
	\While{not converged}{
		Sample LR-HR batch\;
		\While{i$\le$ L}{
			Partiton bands into three groups by Eqs. \ref{group1}-\ref{group3}\;	
			Update $\theta$ by excuting coarse model\;
			i $\leftarrow$ i + 1\;  	
	}}
	Generate coarse model parameters $\theta_c$\;
	Obtain coarse SR results $U$ and $V$ in terms of scale factor $s$\;
	Compute the reconstruction error under spectral angle constrain by Eq. \ref{diff}\;	
	Obtain fine SR result $I_{SR}$ using Eq. \ref{add}.
\end{algorithm}

To extend this algorithm, we further develop back-projection without predefined hyperparameters, which is shown in Fig. \ref{fig:fine_flowchart}.  Specifically, given a LR image $I_{LR}$, the coarse result $U$ is obtained by coarse model $f$ according to scale factor $s$, which is expressed as 
\begin{equation}
U = {\cal F}(s, I_{LR}).
\end{equation}

Previous studies aim to resize SR image to the same size as LR image, and measure the reconstruction error between the two.  As for large scale factor $s$, this approach discards too much useful information, resulting in very little performance gain for reconstruction. Therefore,  in our paper, the  coarse model is utilized to scale LR image $I_{LR}$ again to obtain another initial SR result $V$ for scale factor $\frac{s}{2}$, i.e.,

\begin{equation}
V = {\cal F}(\frac{s}{2}, I_{LR}),
\end{equation}
Then, the coarse result $U$ for scale factor $s$ is  downsampled  using Bicubic.  After generating LR image $M$,  the reconstruction error is reckoned by
\begin{equation}
M = R_{bicu}(0.5, U),
\end{equation} 
\begin{equation}
Z = V - M,
\end{equation}

To  yield high spectral fidelity, we introduce Spectral Angle Mapper (SAM) \cite{kruse1993spectral} to constrain error, which is represented as 
\begin{equation}
\lambda_{SAM} = arccos\left ( \frac{<M, V>}{||M||_2||V||_2}\right ),
\end{equation} 
\begin{equation}
\label{diff}
D{\rm{ = }}\left\{ \begin{array}{l}
{\lambda _{SAM}} \times Z{\kern 1pt},  {\kern 10pt} {\lambda _{SAM}} < 1{\kern 1pt} {\kern 1pt}, \\
Z,{\kern 50pt} otherwise,
\end{array} \right.
\end{equation}
where  $arccos(\cdot)$ is  arccos function.  $<\cdot,\cdot>$  is  dot product. $||\cdot||_2$ denotes the L2 norm. The obtained error is upsampled using Bicubic and added to the coarse result $U$.  Eventually, the fine result $I_{SR}$ is obtained by 

\begin{equation}
N = R_{bicu}(2, D), 
\end{equation}
\begin{equation}
\label{add} 
I_{SR} = U +  N.
\end{equation}

Our method abandons the convolution operation executed in \cite{Yang2010image}. Importantly, the  strategy does not require iteration. Since the refinement is conducted using unsupervised manner, it can be applied to any hyperspectral image SR algorithms to further optimize SR image, which indicates it  belongs to a plug-and-play method. For the description of dual-stage SR algorithm,  the procedures are summarized in \textbf{Algorithm 1}.

\section{Experiments}
In this section, extensive experiments are conducted to evaluate the superiority of the proposed DualSR, and  the comparative results are given from both quantitative and qualitative aspects. 
\label{es}
\subsection{Datasets}
\label{datasets}
The CAVE dataset \cite{yasuma2010generalized} was captured by a tunable filter  and a cooled CCD camera. It contains 32 scenes for a wide variety of objects and materials, which is grouped into 5 parts. The range of full spectral resolution is from 400 $n$m to 700 $n$m at 10 $n$m steps, which means that each scene has 31 bands. The spatial resolution of scene is $512 \times 512$ pixels. Unlike CAVE dataset,  Harvard dataset \cite{chakrabarti2011statistics}  is more challenging. It was collected under daylight, both outdoors and indoors. The dataset contains 71 hyperspectral images. Each scene consists of $1392 \times 1040$ pixels and 31 spectral bands in the range of  420 $n$m to 720 $n$m. In our work, we randomly select 80\% of the data as the training set and the rest for testing. We augment the  given training data by  choosing 24 patches. With respect to each patch, its size is scaled 1, 0.75, and 0.5 times, respectively. We rotate these patches  $90^{\circ}$ and  flip them horizontally. Through various blur kernels, we then subsample these patches into LR hyperspectral images with the size of $L \times 32 \times 32$.

\subsection{Implementation Details}
For our network, the convolution kernel after concatenation is set to $1 \times 1$, which reduces the number of channels.  We adopt sub-pixel convolution layer to upscale the features into HR space in terms of upsampling  operation. The kernel of other convolution operations involved in the network is fixed to $3 \times 3$, and the number of  convolution kernels is defined as 64. In the training phase, our network is trained using $L1$ loss function. The mini-batch is set to 64. We optimize our network using ADAM optimizer with $\beta_1$ = 0.9 and $\beta_2$ = 0.999 and initial learning rate $10^{-4}$. For learning rate, it  is gradually updated  by a half at every 30 epochs. Our model runs on PyTorch framework and is trained with NVIDIA GTX 1080 GPU.

\subsection{Evaluation Metrics}
To quantitatively evaluate  the proposed method, we apply Peak Signal-to-Noise Ratio (PSNR), Structural SIMilarity (SSIM), and Spectral Angle Mapper (SAM). Among these metrics, PSNR and SSIM are to evaluate the performance of super-resolved hyperspectral image in spatial domain. Generally, the higher their values  are, the better the performance is. SAM is to analyze the performance of restored image in spectral domain. The smaller the value is, the less the spectral distortion is. 

\subsection{Ablation Study}
\begin{table}[]
	\centering
	\renewcommand\arraystretch{1.1}
	\caption{Ablation study for scale factor $\times 4$ on CAVE dataset. Bicubic SR kernel is used.}
	\label{table:AS}
	\begin{tabular}{c|cccc}
		\toprule[0.8pt]
		Module             & \multicolumn{4}{c}{Different combinations of modules} \\ \hline
		Inter-group fusion & \ding{52}   & \ding{55}   & \ding{52}& \ding{52}  \\
		Intra-group fusion & \ding{55}  & \ding{52}   & \ding{52}  & \ding{52}   \\
		FCF                & \ding{55}  & \ding{55}   & \ding{55} & \ding{52}   \\ \hline
		PSNR               & 38.720     &  39.521     & 39.635   & 39.665     \\
		SSIM               & 0.9295    &  0.9322   & 0.9327   & 0.9328     \\
		SAM                & 3.304      & 3.211      & 3.151     & 3.152        \\ \toprule[0.8pt]
	\end{tabular}
\end{table}
To verify the effectiveness of key modules, we test variant versions of our model by removing each component, including  inter-group fusion, intra-group fusion, and FCF. 

Table \ref{table:AS} shows the ablation study for key modules. For clear show, we only provide the results evaluated by CoarSR. Specifically, with intra-group fusion, the network directly exploits 2D convolution to explore spatial information. In contrast with only the existence of inter-group fusion, the network exhibits  better performance.  It implies that the model should pay more attention to the exploration of spatial features and contribute to generating sharper results, which is consistent with the original intention of our design. When two kinds of group fusion are attached, the network tends to bring fine results. It indicates the combination of the two effectively integrates complementary information between different groups. 

Finally, all three modules are involved in feature extraction. One can observe that this combination produces better performance than any other arbitrary module combination, except for SAM. It is concluded that these quantitative analyses revels the effectiveness and benefits of the proposed modules.

\subsection{Study of Enhanced Back-projection} To further refine result obtained by CoarSR, we propose an enhanced back-projection under spectral angle constraint. In this section, we study  the effectiveness of enhanced back-projection on bicubic SR kernel. Note that the proposed enhanced back-projection involves two different scale factors during  fine stage. Therefore, the results are evaluated for scale factors $\times 4$ and $\times 8$ on two public datasets.

\subsubsection{\textbf{Performance Comparison of Back-projection}}
Since the proposed enhanced back-projection belongs to the post-processing method and is unsupervised, it is suitable for any hyperspectral image algorithms to further boost the performance. To investigate the performance gain for different SR algorithms, we apply it as post-processing of some methods to analyze and discuss. When the scale factor is set to 4, the designed back-projection method involves two different scale factors, i.e., $\times 2$ and $\times 4$. In SSPSR, the minimum upscale factor is set to $\times4$ due to  progressive upsampling. Therefore, the results of scale factor $\times 4$  cannot be provided when conducting our designed back-projection.

Table \ref{tab:fine} shows the differences of two scale factors with or without post-processing. One can notice that all methods markedly reveal performance gains under action of enhanced back-projection, compared with without post-processing. Moreover, we also present the performance by setting  back-projection from \cite{Yang2010image} as post-processing. Different from our enhanced  back-projection, \cite{Yang2010image} only contains one kind of scale factor. Hence, we provide the results obtained by SSPSR for scale factor $\times 4$.   It can be seen in Table \ref{tab:otherfine} that although the method can produce relatively good performance, it is not as obvious as the improvement in our work. Notably, the spectral distortion actually gets worse. Through analysis, this is enough to prove the effectiveness of our method.

\begin{table*}[]
	\centering
	\renewcommand\arraystretch{1.2}	
    \caption{Performance gain  by setting  enhanced back-projection method as post-processing for scale factor $\times 4$ on CAVE dataset. }
	\label{tab:fine}
	\begin{tabular}{c|c|c|cccc|cccc}
		\toprule[0.8pt]
		\multicolumn{1}{c|}{\multirow{2}{*}{\begin{tabular}[c]{@{}c@{}}Post- \\ processing\end{tabular}}} & \multirow{2}{*}{Scale}      & \multirow{2}{*}{Metrics} & \multicolumn{4}{c|}{CAVE}         & \multicolumn{4}{c}{Harvard}       \\ \cline{4-11} 
		\multicolumn{1}{c|}{}                                 &                             &                         & SSPSR \cite{9097432} & SFCSR \cite{Wang2020sfcsr} & ERCSR  \cite{li2020exp}  & CoarSR & SSPSR\cite{9097432}  & SFCSR \cite{Wang2020sfcsr}  & ERCSR  \cite{li2020exp} & CoarSR\\ \hline
		\multirow{6}{*}{\ding{55}}      & \multirow{3}{*}{$\times 4$} & PSNR                     & 38.366 & 39.192 & 39.224 & 39.665 & 39.293 & 40.077 & 40.211 & 40.312 \\
		&                             & SSIM                     & 0.9272 & 0.9321 & 0.9322 & 0.9328 & 0.9333 & 0.9373 & 0.9374 & 0.9405 \\
		&                             & SAM                      & 3.484  & 3.221  & 3.243  & 3.152  & 2.448  & 2.407  & 2.384  & 2.384  \\ \cline{2-11} 
		& \multirow{3}{*}{$\times 8$} & PSNR                     & 34.151 & 35.294 & 35.102 & 35.475 & 34.435 & 35.097 & 35.252 & 35.628 \\
		&                             & SSIM                     & 0.8718 & 0.8828 & 0.8824 & 0.8828 & 0.8676 & 0.8737 & 0.8733 & 0.8755 \\
		&                             & SAM                      & 4.831  & 4.378  & 4.445  & 4.269 & 2.899  & 2.911  & 2.888  & 2.794\\ \hline
		\multirow{6}{*}{\ding{52}}      & \multirow{3}{*}{$\times 4$} & PSNR                     & --      & 39.372 & 39.334 & 39.813 & --      & 40.257 & 40.308 & 40.540  \\
		&                             & SSIM                     & --      & 0.9335 & 0.9334 & 0.9339 & --      & 0.9394 & 0.9401 & 0.9423 \\
		&                             & SAM                      & --      & 3.201  & 3.196  & 3.130   & --      & 2.375  & 2.367  & 2.372  \\ \cline{2-11} 
		& \multirow{3}{*}{$\times 8$} & PSNR                     & 34.453 & 35.341 & 35.479 & 35.628 & 34.855 & 35.414 & 35.564 & 35.628 \\
		&                             & SSIM                     & 0.8771 & 0.8846 & 0.8855 & 0.8862 & 0.8719 & 0.8775 & 0.8793 & 0.8799 \\
		&                             & SAM                      & 4.661  & 4.267  & 4.264  & 4.141 & 2.837  & 2.803  & 2.829  & 2.774  \\ \toprule[0.8pt]
	\end{tabular}
\end{table*}


\begin{table*}[]
	\centering
	\renewcommand\arraystretch{1.2}	
	\caption{Performance gain  by setting  \cite{Yang2010image} as post-processing for scale factor $\times 4$ on CAVE dataset. }
	\label{tab:otherfine}
	\begin{tabular}{c|c|cccc|cccc}
		\toprule[0.8pt]
		\multirow{2}{*}{Scale}      & \multirow{2}{*}{Metrics} & \multicolumn{4}{c|}{CAVE}         & \multicolumn{4}{c}{Harvard}       \\ \cline{3-10} 
		&                           & SSPSR \cite{9097432} & SFCSR \cite{Wang2020sfcsr} & ERCSR  \cite{li2020exp}  & CoarSR  & SSPSR\cite{9097432}  & SFCSR \cite{Wang2020sfcsr}  & ERCSR  \cite{li2020exp} & CoarSR \\ \hline
		\multirow{3}{*}{$\times 4$} & PSNR     & 38.426 & 39.225 & 39.244 & 39.675 & 39.350  & 40.089 & 40.224 & 40.327 \\
		& SSIM                     & 0.9286 & 0.9326  & 0.9330  & 0.9332 & 0.9339 & 0.9380  & 0.9395 & 0.9406 \\
		& SAM                      & 3.545  & 3.230  & 3.229  & 3.165  & 2.434  & 2.406  & 2.382  & 2.383  \\ \hline
		\multirow{3}{*}{$\times 8$} & PSNR                     & 34.335 & 35.335 & 35.147 & 35.497 & 34.538 & 35.118 & 35.179 & 35.270  \\
		& SSIM                     & 0.8730  & 0.8835 & 0.8830  & 0.8832 & 0.8680  & 0.8737 & 0.8735 & 0.8756 \\
		& SAM                      & 4.922  & 4.398  & 4.467  & 4.283  & 2.905  & 2.925  & 2.858  & 2.814  \\ \toprule[0.8pt]
	\end{tabular}
\end{table*}

\subsubsection{\textbf{Effect of Spectral Angle Constraint}}

In our work, we introduce Spectral Angle Mapper (SAM) to constrain reconstruction error, which can alleviate spectral distortion. To explore whether the constraint is effective, we delete it and see how the performance changes. Table \ref{tab:asc} depicts  the results  using enhanced back-projection without spectral angle constraint.  We combine the results in Table \ref{tab:fine} for analysis. As one can see, some values do not change because the value calculated for SAM is greater than 1. As a result, the constraint does not work via Eq. \ref{diff}.   In addition to that,  we observe that spectral angle constraint does enhance spectral fidelity  in most cases. Meanwhile, the performance of other metrics has also been improved to a certain extent.  It indicates our method can effectively integrate spatial information under constraint and make the spectral curves at different scales as consistent as possible.

\begin{table*}[]
	\centering
	\renewcommand\arraystretch{1.2}	
	\caption{Performance gain  by setting enhanced Back-projection without spectral angle constraint as post-processing for scale factor $\times 4$ on CAVE dataset. }
	\label{tab:asc}
	\begin{tabular}{c|c|cccc|cccc}
		\toprule[0.8pt]
		\multirow{2}{*}{Scale}      & \multirow{2}{*}{Metrics} & \multicolumn{4}{c|}{CAVE}         & \multicolumn{4}{c}{Harvard}       \\ \cline{3-10} 
		&                           & SSPSR \cite{9097432} & SFCSR \cite{Wang2020sfcsr} & ERCSR  \cite{li2020exp}  & CoarSR  & SSPSR\cite{9097432}  & SFCSR \cite{Wang2020sfcsr}  & ERCSR  \cite{li2020exp} & CoarSR  \\ \hline
		\multirow{3}{*}{$\times 4$} & PSNR    & --     & 39.221 & 39.317 & 39.774 & --       & 40.246 & 40.295 & 40.541 \\
		& SSIM                     & --       & 0.9330  & 0.9333 & 0.9338 & --       & 0.9394 & 0.9396 & 0.9422 \\
		& SAM                      & --       & 3.205  & 3.209  & 3.141  & --       & 2.381  & 2.376  & 2.385  \\ \hline
		\multirow{3}{*}{$\times 8$} & PSNR                     & 34.453 & 35.341 & 35.479 & 35.609 & 34.928 & 35.417 & 35.574 & 35.675 \\
		& SSIM                     & 0.8771 & 0.8846 & 0.8855 & 0.8861 & 0.8724 & 0.8774 & 0.8796 & 0.8799 \\
		& SAM                      & 4.661  & 4.267  & 4.264  & 4.145  & 2.839  & 2.805  & 2.837  & 2.786  \\ \toprule[0.8pt]
	\end{tabular}
\end{table*}

\subsection{Generalizability to Various Datasets}

Using known bicubic downsampling condition, we compare our proposed DualSR  with existing multiple approaches on CAVE and Harvard datasets, including 3D-FCNN \cite{Mei2017Hyperspectral}, EDSR \cite{Lim2017Enhanced}, SSPSR \cite{9097432}, MCNet \cite{Li2020Mixed}, SFCSR \cite{Li2020Mixed}, ERCSR \cite{li2020exp}. Next, we investigate the performance by quantitative evaluation and qualitative comparison.
\begin{table*}[tph]
	\centering
	\renewcommand\arraystretch{1.1}	
	\caption{Quantitative evaluation of existing SR approaches for different scale factors on CAVE dataset. The best results are {\color{red}red} font, and the second best results are {\color{blue}blue} font.}
	\label{tab:cave}
	\begin{tabular}{c|c|ccccccccc}
		\toprule[0.8pt]
        Scale      & Metric & Bicubic & 3D-FCNN \cite{Mei2017Hyperspectral} & EDSR \cite{Lim2017Enhanced}  & SSPSR \cite{9097432}  & MCNet \cite{Li2020Mixed} & SFCSR \cite{Wang2020sfcsr} & ERCSR \cite{li2020exp} & \begin{tabular}[c]{@{}c@{}}\textbf{CoarSR} \\ (\textbf{Ours})\end{tabular}  & \begin{tabular}[c]{@{}c@{}}\textbf{DualSR} \\ (\textbf{Ours})\end{tabular} \\ \hline
		\multirow{3}{*}{$\times 4$} & PSNR $\uparrow$   & 35.755  & 37.626  & 38.587 & 38.366 & 39.026 & 39.192 & 39.224 &{\color{blue}39.665}      &{\color{red}39.813}        \\
		& SSIM $\uparrow$   & 0.9071  & 0.9195  & 0.9292 & 0.9272 & 0.9319 & 0.9321 & 0.9322 &{\color{blue}0.9328}      &{\color{red}0.9339}        \\
		& SAM $\downarrow$   & 3.944   & 3.360   & 3.804  & 3.484  & 3.292 & 3.221  & 3.243  &{\color{blue}3.152}      & {\color{red}3.130}       \\ \hline
		\multirow{3}{*}{$\times 8$} & PSNR $\uparrow$   & 31.805  & 32.956  & 31.554 & 34.151 & 35.320 & 35.294 & 35.102 &{\color{blue}35.475}      &  {\color{red}35.628}      \\
		& SSIM $\uparrow$   & 0.8485  & 0.8600  & 0.8233 & 0.8718 & {\color{blue}0.8833} & 0.8828 & 0.8824 &0.8828      &  {\color{red}0.8862}      \\
		& SAM $\downarrow$   & 5.291   & 4.384   & 10.670 & 4.831 & 4.423 & 4.378  & 4.445  &{\color{blue}4.269}      &  {\color{red}4.4107}      \\ \toprule[0.8pt]
	\end{tabular}
\end{table*}
\begin{figure}[thp]
	\centering
	\includegraphics[height= 4.9cm, width=0.39\textwidth]{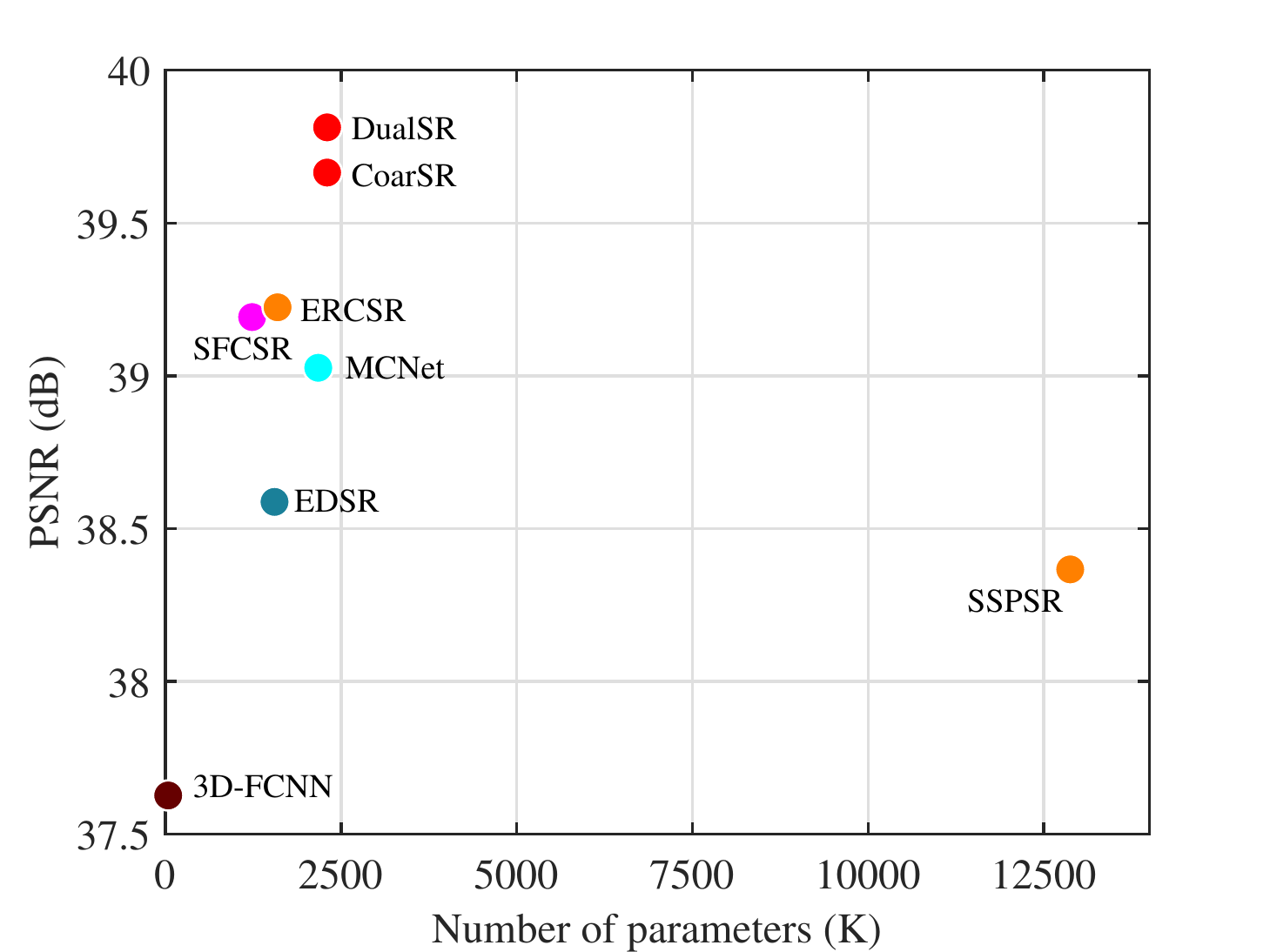}
	\caption{PSNR performance versus number of parameters. The results between both are provided for scale factor $\times 4$ on CAVE dataset. Our CoarSR and DualSR obtain better trade off between performance and model size.}
	\label{fig:para}
\end{figure}
\begin{table*}[tp]
	\centering
	\renewcommand\arraystretch{1.1}	
	\caption{Quantitative evaluation of existing SR approaches for different scale factors on Harvard dataset. The best results are {\color{red}red} font, and the second best results are {\color{blue}blue} font.}
	\label{table:harvard} 	
	\begin{tabular}{c|c|ccccccccc}
		\toprule[0.8pt]
		Scale      & Metric & Bicubic & 3D-FCNN \cite{Mei2017Hyperspectral} & EDSR \cite{Lim2017Enhanced}  & SSPSR \cite{9097432} &MCNet \cite{Li2020Mixed}  & SFCSR \cite{Wang2020sfcsr} & ERCSR \cite{li2020exp} &  \begin{tabular}[c]{@{}c@{}}\textbf{CoarSR} \\ (\textbf{Ours})\end{tabular}  & \begin{tabular}[c]{@{}c@{}}\textbf{DualSR} \\ (\textbf{Ours})\end{tabular} \\ \hline
		\multirow{3}{*}{$\times 4$} & PSNR $\uparrow$   & 37.227  & 38.143  & 39.175 & 39.293 &40.081 &40.077 & 40.211 &{\color{blue}40.312}      &{\color{red}40.540}       \\
		& SSIM $\uparrow$  & 0.9122  & 0.9188  & 0.9324 & 0.9333 & 0.9367 & 0.9373 & 0.9374 &{\color{blue}0.9405}     & {\color{red}0.9423}       \\
		& SAM  $\downarrow$   & 2.531   & 2.363   & 2.560  & 2.448  & 2.410 &2.407 & {\color{blue}2.384}  &{\color{blue}2.384}      &{\color{red}2.372}        \\ \hline
		\multirow{3}{*}{$\times 8$} & PSNR $\uparrow$   & 33.275  & 33.363  & 34.068 & 34.435 &34.927 & 35.097 & 35.138 &{\color{blue}35.252}      & {\color{red}35.628}       \\
		& SSIM $\uparrow$  & 0.8518  & 0.8448  & 0.8598 & 0.8676 &0.8732 & 0.8737 & 0.8733 &{\color{blue}0.8755}      & {\color{red}0.8799}       \\
		& SAM  $\downarrow$   & 2.884   & 2.726   & 3.051 & 2.899 &2.957 & 2.911  & 2.888  &{\color{blue}2.794}      & {\color{red}2.774}       \\ \toprule[0.8pt]
	\end{tabular}
\end{table*}


\subsubsection{\textbf{Quantitative Evaluation}} Tables \ref{tab:cave} and \ref{table:harvard} show the quantitative evaluation of existing SR approaches for different scale factors on two datasets. We can notice that our method attains excellent results regardless of whether there is fine stage. Among the approaches only using 2D convolution, SSPSR attains superior performance across datasets. It first fuses features between groups for small scale factor, and then performs SR for large scale factor. The progressive upsampling is more conducive to the generation of clear image. Thanks to the exploration of spectral information, in contrast, the model using 3D convolution has better performance than the model using 2D convolution on the whole, except for 3D-FCNN. With respect to 3D-FCNN,  the main reasons for its low results are the shallow design and the lack of residual learning. With respect to recurrent networks, i.e., SFCSR and CoarSR,  they exploit spectral structure from multiple neighboring bands to achieve complementary information. Their structures are more helpful to generate a single clear band. Although the number of parameters of CoarSR is more than that of SFCSR (see Fig. \ref{fig:para}), the results obtained by our CoarSR are significantly higher than that of SFCSR, i.e., +0.473 dB, 0.0007, -0.069 on scale factor $\times 4$ on CAVE dataset.  In our opinion, the main reasons are in two aspects.  The one is that more representation learning is added in spatial domain, when spectral information is available. The other is that our method fully exploit the complementary information of several neighboring bands, recovering  missing details. Interestingly, without increasing the number of parameters, the values of proposed DualSR increase obviously in all three metrics, outperforming others by a non-negligible margin.

\subsubsection{\textbf{Qualitative Evaluation}}
\label{bicu}
\begin{figure*}[tp]
	\centering
	\includegraphics[height= 7.8cm, width=0.95\textwidth]{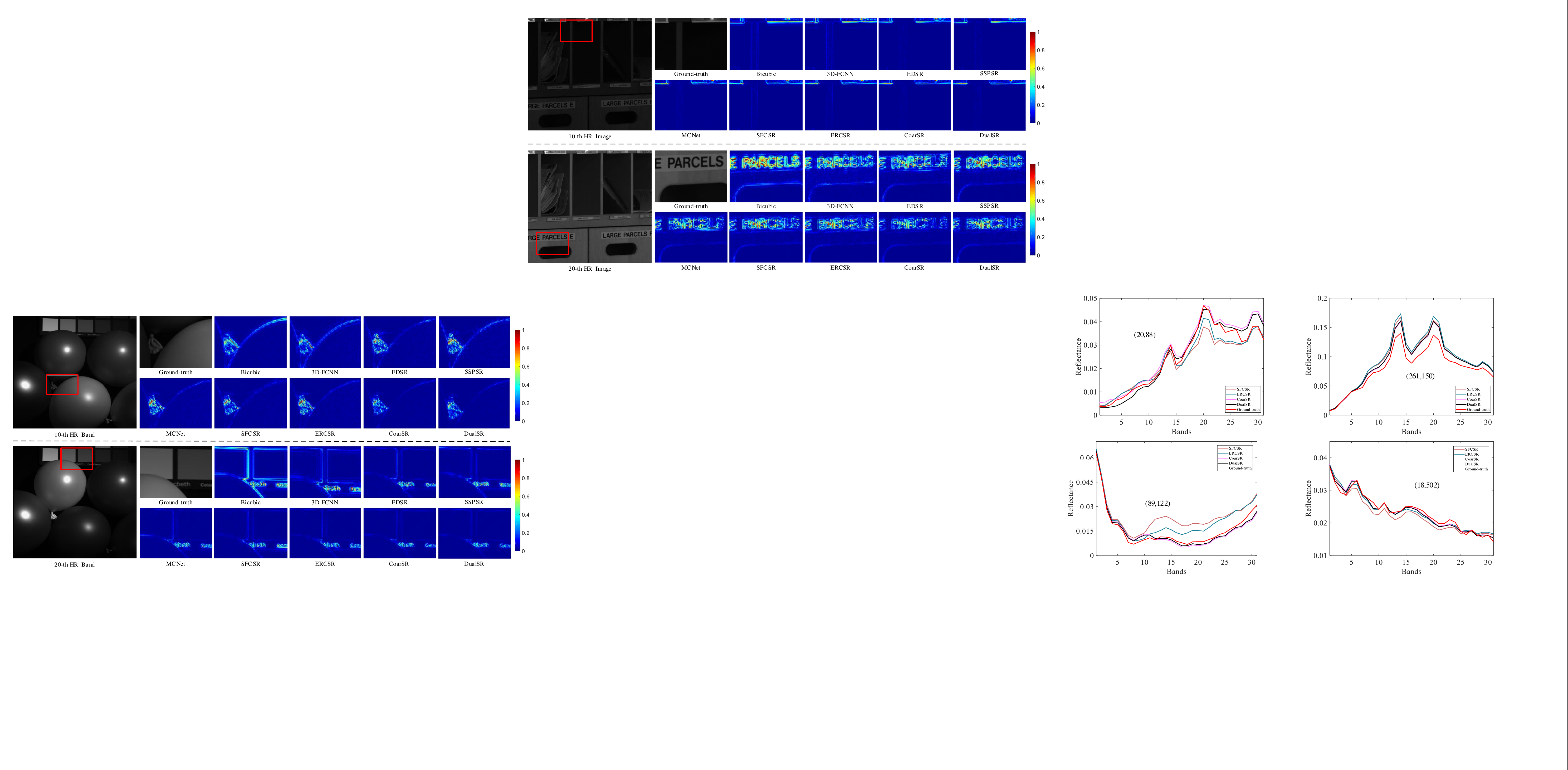}
	\caption{Visual results in terms of spatial domain with existing SR methods on CAVE dataset. The results  of  \textit{balloons} image are evaluated for scale factor $\times 4$. The first line denotes SR results of 10\textit{-th} band, and the second line denotes SR results of 20\textit{-th} band.}
	\label{fig:cave}
\end{figure*}

\begin{figure*}[tp]
	\centering
	\includegraphics[height= 7.8cm, width=0.95\textwidth]{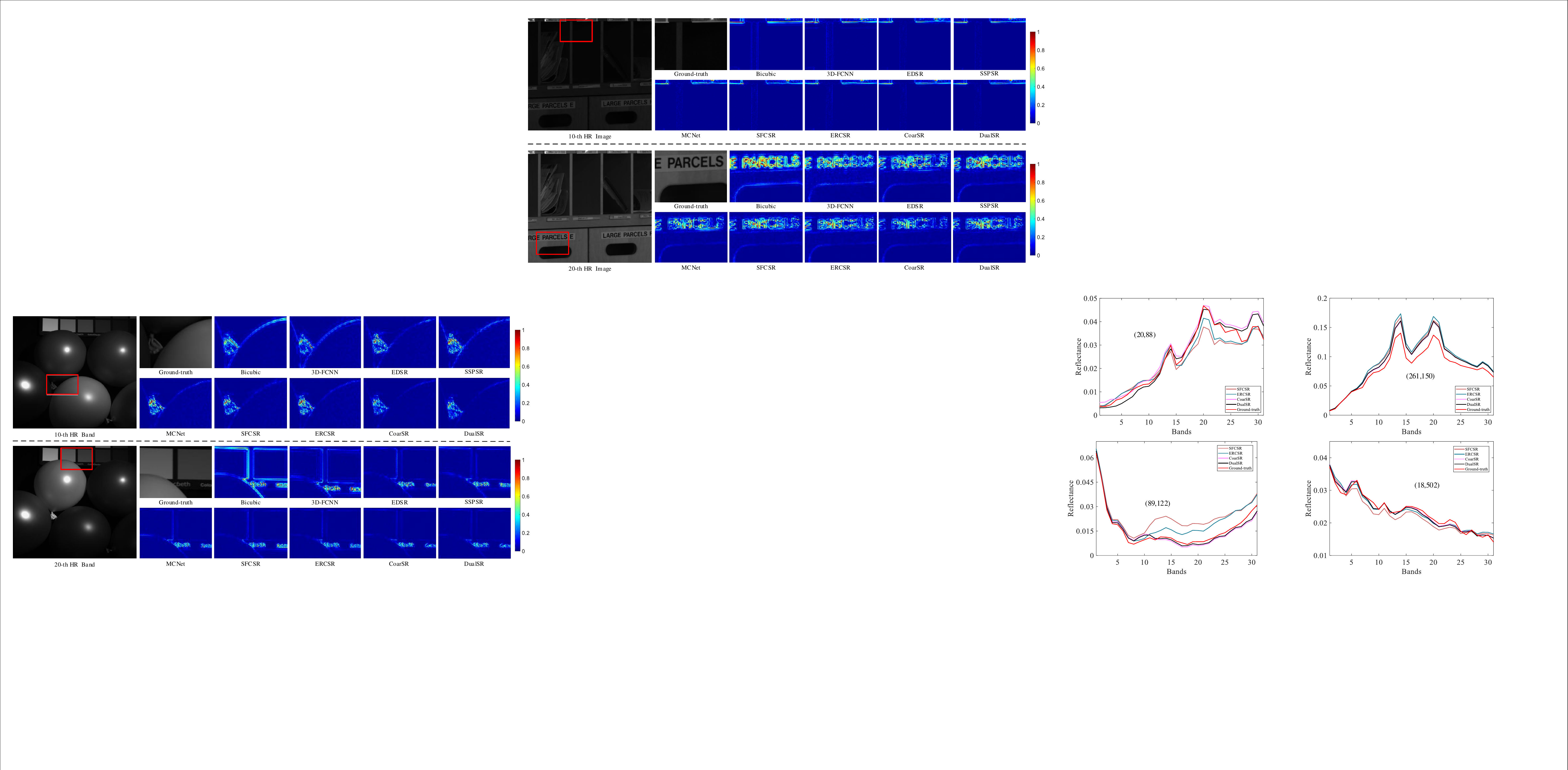}
	\caption{Visual results in terms of spatial domain with existing SR methods on Harvard dataset. The results of \textit{imgd5} image are evaluated for scale factor $\times 4$. The first line denotes SR results of 10\textit{-th} band, and the second line denotes SR results of 20\textit{-th} band.}
	\label{fig:harvard}
\end{figure*}

We further analyze the proposed method by qualitative evaluation.  To show the visual results in terms of spatial domain, in our work, two bands in single SR hyperspectral image are provided, i.e., 10\textit{-th} band and 20\textit{-th} band. Since the band is gray, it does not effectively observe the edge details of image. To clearly distinguish the difference  between super-resolved and HR image, the  absolute error map is given. Figs. \ref{fig:cave} and \ref{fig:harvard} present visual example for scale factor $\times$4 on two datasets.  One observe that our method produces low absolute errors. In  particular, there are more shallow edges in some positions, which  indicates that the proposed approach can generate sharper edges and finer details.  It  is consistent with the analysis in Tables \ref{tab:cave} and \ref{table:harvard}, which further demonstrates  that our approach can simultaneously learn spectral and spatial knowledge while generating diverse textures.
\begin{figure*}[thp]
	\centering
	\includegraphics[height= 3.3cm, width=1\textwidth]{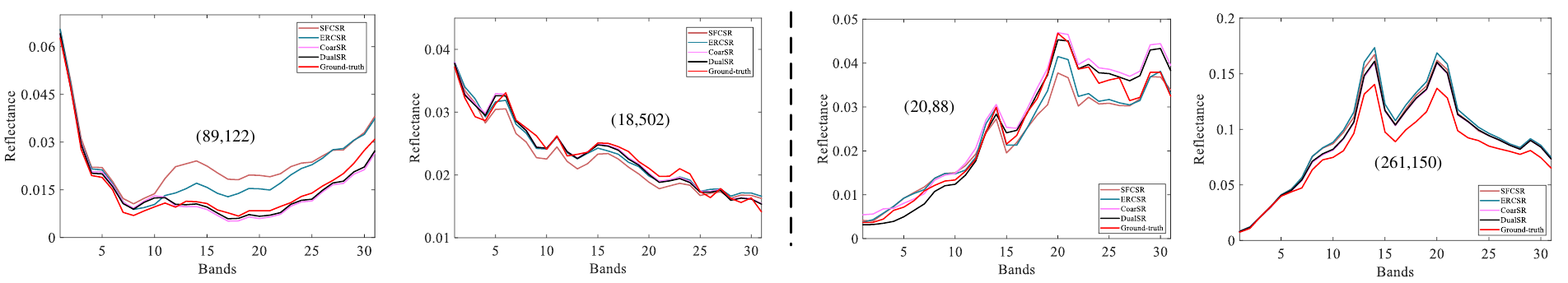}
	\caption{Visual comparison  in terms of spectral domain by randomly selecting two pixels  for scale factor $\times 4$. The two  on the left are the results of \textit{balloons} image on CAVE dataset.  The two  on the right are the results of \textit{imgd5} image on Harvard dataset.   Note that  to avoid confusion, only two representative algorithms are compared with our methods.}
	\label{fig:spectral}
\end{figure*}

We also exhibit the visual results in spectral domain. Concretely, one pixel in  spatial domain is selected, and the spectral curve is plotted along spectral dimension at the same position. Fig.  \ref{fig:spectral}   displays the spectral distortion of super-resolved hyperspectral image by randomly choosing two pixels on two datasets, respectively.  We can see that our  DualSR  maintains the same curve as the ground-truth in most cases.  It validates that the proposed method  can yield higher spectral fidelity against other approaches.

\subsection{Generalizability to Various Blur Kernels}
\begin{table*}[tph]
	\centering
	\renewcommand\arraystretch{1.1}	
	\caption{SR in the presence of various downscaling  kernels. The best results are {\color{red}red} font, and the second best results are {\color{blue}blue} font.}
	\label{tab:vbk}
	\begin{tabular}{c|c|ccccccccc}
		\toprule[0.8pt]
		Scale      & Metric & Bicubic & 3D-FCNN \cite{Mei2017Hyperspectral} & EDSR \cite{Lim2017Enhanced}  & SSPSR \cite{9097432}  & MCNet \cite{Li2020Mixed} & SFCSR \cite{Wang2020sfcsr} & ERCSR \cite{li2020exp} & \begin{tabular}[c]{@{}c@{}}\textbf{CoarSR} \\ (\textbf{Ours})\end{tabular}  & \begin{tabular}[c]{@{}c@{}}\textbf{DualSR} \\ (\textbf{Ours})\end{tabular} \\ \hline
		\multirow{3}{*}{$\times 4$} & PSNR $\uparrow$   & 35.652 &37.263   & 38.204 & 37.932 & 38.317 & 38.410 & 38.390 & {\color{blue}39.296}     &  {\color{red}39.432}     \\
		& SSIM $\uparrow$   & 0.9067  &0.9175   &0.9263  & 0.9253 & 0.9281 & 0.9281 & 0.9286 &{\color{blue}0.9309}      & {\color{red} 0.9312}      \\
		& SAM $\downarrow$   & 3.871   &3.241   &3.601  & 3.578  & 3.410 & 3.390  & 3.403  &{\color{blue}3.242}      & {\color{red}3.238}       \\ \hline
		\multirow{3}{*}{$\times 8$} & PSNR $\uparrow$   & 31.735  &32.813  &34.250  & 33.920 &34.601  & 34.635 & 34.503 &{\color{blue} 34.972}   &  {\color{red}35.322}     \\
		& SSIM $\uparrow$   & 0.8496  & 0.8584  &0.8711  & 0.8699 &0.8787  & 0.8828 & 0.8784 & {\color{blue}0.8802}    &  {\color{red}0.8842}     \\
		& SAM $\downarrow$   & 5.134   &4.302   &4.855  & 4.893 &4.638  & 4.893  & 4.727  & {\color{blue}4.462}    & {\color{red}4.256}     \\ \toprule[0.8pt]
	\end{tabular}
\end{table*}
In this section, we show  results on various blur kernel conditions to demonstrate generalizability. For fair comparison, each image is subsampled  by different downsampling kernels using \cite{Michaeli2013}. The LR images are input into the model to get the corresponding restored images, and the numerical results are obtained by calculating the average value. Note that the downscaling kernels in our work are  \textit{cubic}, \textit{lanczos}, \textit{box}, and \textit{linear}. 
\begin{figure*}[tp]
	\centering
	\includegraphics[height= 7.8cm, width=0.95\textwidth]{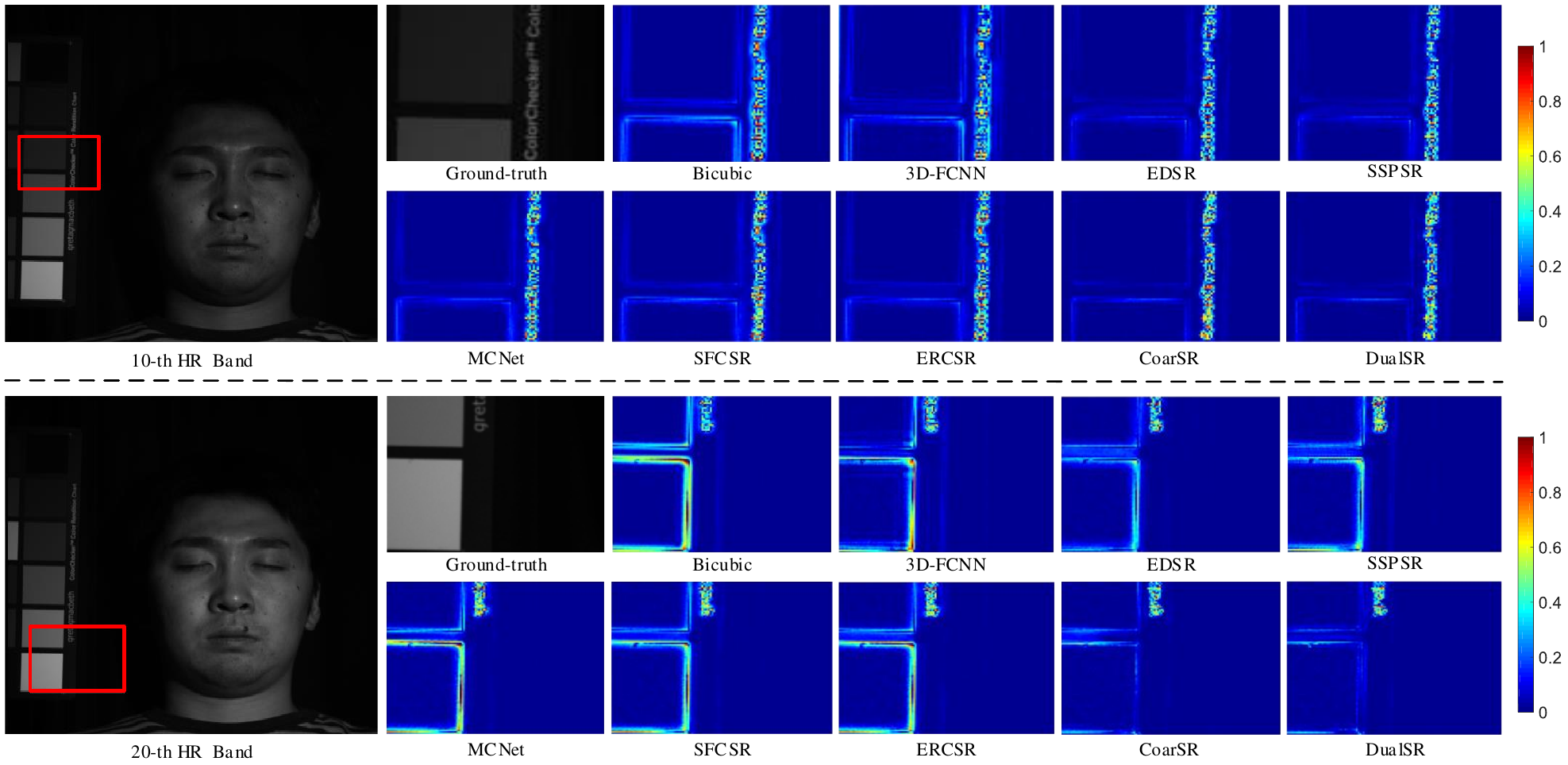}
	\caption{Visual results subsampled by linear kernel  in terms of spatial domain on CAVE dataset. The results  of  \textit{face} image are evaluated for scale factor $\times 4$. The first line denotes SR results of 10\textit{-th} band, and the second line denotes SR results of 20\textit{-th} band.}
	\label{fig:cave_mk}
\end{figure*}

\begin{figure}[tp]
	\centering
	\includegraphics[height= 3.3cm, width=0.48\textwidth]{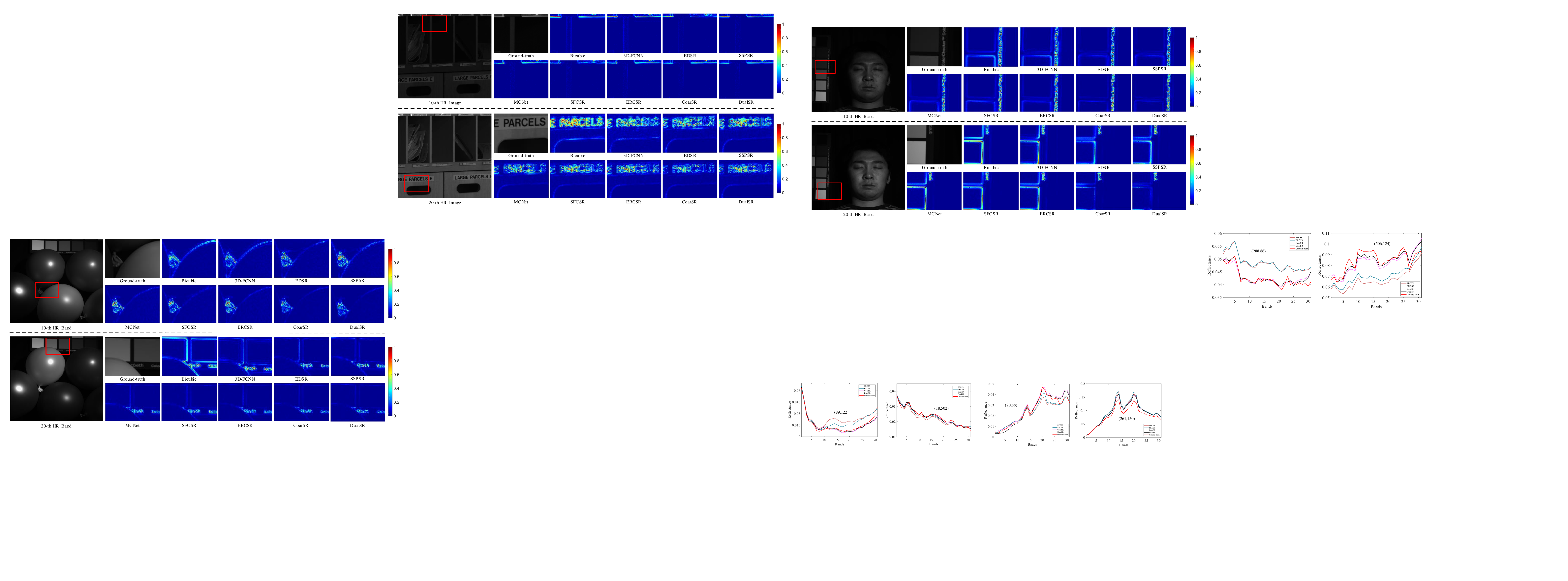}
	\caption{Visual comparison  in terms of spectral domain by randomly selecting two pixels   for scale factor $\times 4$.  Note that  to avoid confusion, only two representative algorithms are compared with our methods.}
	\label{fig:cave_mk_s}
\end{figure}
\subsubsection{\textbf{Quantitative Evaluation}}Table \ref{tab:vbk} compares our performance against other methods on CAVE dataset. As  can be seen, these conventional SR networks obtain severe performance drops due to fixed bicubic SR kernel, which deviates from the non-bicubic SR kernels. It indicates these methods cannot deal with other non-bicubic SR kernels.  On the contrary, we observe that the proposed CoarSR can address non-bicubic SR kernel degraded images well, and its performance  exceeds that of  SFCSR for scale factor $\times 4$ by a very large margin: + 1.022 dB.  This is due to the fact that our method uses various downscaling  kernels in the process of training, while other methods do not. From this perspective,  CoarSR has better generalization performance, compared to other methods.  In other words, the proposed approach is more robust across different scenarios. 

\subsubsection{\textbf{Qualitative Evaluation}}
Similarly, we take the same approach from Section \ref{bicu} to provide the visual results in terms of spatial  and spectral domain. To make a simple comparison, the results subsampled by linear downscaling kernel are exhibited, which is shown in Figs. \ref{fig:cave_mk} and \ref{fig:cave_mk_s}. As seen, our method still achieves better performance in two aspects, which reveals it can deal with LR image subsampled by  non-bicubic kernel well.  Through the above analysis, it proves the proposed model surpasses  state-of-the-art networks over each dimension on various blur kernel conditions.

\section{Conclusion}
\label{con}
In this study, we propose a new structure for hyperspectral image SR, which contains coarse stage and fine stage. In coarse stage,  the coarse SR image is obtained in band-by-band. Different from previous works, in this process, we add more adjacent bands to the model, and encourage the network by grouping them to learn the potential information of the current band. With the intra/inter-group fusion, the complementary information is borrowed from adjacent bands, achieving refined details. To learn the content of spectal-spatial consistency,  an enhanced back-projection method is proposed.  This method can be utilized as the post-processing for any hyperspectral image SR algorithms, dramatically boosting performance gain. Experiments demonstrate our proposed DualSR can produce state-of-the-art results over existing works. Compared with second best method, the designed approach in PSNR  and SSIM increase by 0.589dB and 0.0017 for scale factor $\times 4$, and its SAM decreased by 0.113.

In the future, we  plan to extend our method by constructing more realistic training pairs. In the real scene, the image degradation process is complicated due to noise motion blur and other factors, and the performance of the traditional SR network which only uses Bicubic kernel training decreases obviously in the face of other kernels. Thus, our next focus is to build more realistic training data.

\bibliographystyle{IEEEbib}
\bibliography{refs}	

\end{document}